\documentclass[prl,twocolumn,notitlepage,superscriptaddress,longbibliography]{revtex4-1}
\usepackage{amsmath}
\usepackage{amssymb}

\usepackage[unicode=true,colorlinks=true,citecolor=blue,urlcolor=blue]{hyperref}

\usepackage{bm}
\usepackage{epsfig}

\usepackage[normalem]{ulem}

\newcommand {\diag}{\mathop{\mathrm{diag}}\nolimits}

\renewcommand {\Im}{\mathop\mathrm{Im}\nolimits}

\renewcommand {\phi}{{\varphi}}
\newcommand {\rmi}{{\rm i}}
\newcommand {\rmd}{{\rm d}}

\newcommand {\e}{{\rm e}}
\newcommand {\eps}{\varepsilon}

\begin{document}
\title{Photon-mediated localization in two-level qubit arrays
}
\author{Janet Zhong}
\affiliation{Nonlinear Physics Centre, Research School of Physics, Australian National University, Canberra ACT 2601, Australia}
\author{Nikita A. Olekhno}
\affiliation{ITMO University, St. Petersburg 197101, Russia}
\author{Yongguan Ke}
\affiliation{Nonlinear Physics Centre, Research School of Physics, Australian National University, Canberra ACT 2601, Australia}
\affiliation{Laboratory of Quantum Engineering and Quantum Metrology, School of Physics and Astronomy,
Sun Yat-Sen University (Zhuhai Campus), Zhuhai 519082, China}
\author{Alexander V. Poshakinskiy}
\affiliation{Ioffe Institute, St. Petersburg 194021, Russia}
\author{Chaohong Lee}
%\email{lichaoh2@mail.sysu.edu.cn}
\affiliation{Laboratory of Quantum Engineering and Quantum Metrology, School of Physics and Astronomy,
Sun Yat-Sen University (Zhuhai Campus), Zhuhai 519082, China}
\affiliation{State Key Laboratory of Optoelectronic Materials and Technologies, Sun Yat-Sen University (Guangzhou Campus), Guangzhou 510275, China}

\author{Yuri S. Kivshar}
\affiliation{Nonlinear Physics Centre, Research School of Physics, Australian National University, Canberra ACT 2601, Australia}
\affiliation{ITMO University, St. Petersburg 197101, Russia}
\author{Alexander N. Poddubny}
\email{poddubny@coherent.ioffe.ru}
\affiliation{Nonlinear Physics Centre, Research School of Physics, Australian National University, Canberra ACT 2601, Australia}
\affiliation{ITMO University, St. Petersburg 197101, Russia}
\affiliation{Ioffe Institute, St. Petersburg 194021, Russia}

\begin{abstract}
We predict the existence of a novel interaction-induced spatial localization in a periodic array of qubits coupled to a waveguide. 
This localization can be described as a quantum analogue of a self-induced optical lattice between two indistinguishable photons, where one photon creates a standing wave that traps the other photon.  The localization is caused by the interplay between on-site repulsion due to the photon blockade and the waveguide-mediated long-range coupling between the qubits.
\end{abstract}
\date{\today}

\maketitle
{\it Introduction.} Many-body quantum optical systems have received intense interest in the recent years due to ground-breaking experiments with superconducting qubits~\cite{vanLoo2013,Mirhosseini2019,Wang2019} and cold atoms coupled to waveguides~\cite{Corzo2019}. A paradigmatic system in quantum optics is an array of atoms coupled to freely propagating photons~\cite{Dicke1954,Roy2017,KimbleRMP2018}. 
Waveguide quantum electrodynamics, where photons propagate in one dimension, is promising for quantum networks~\cite{kimble2008quantum} and quantum computation~\cite{Zheng2013}. When atoms are located at the same point, the  full quantum problem can be solved exactly~\cite{Yudson1984} because light interacts only with the  symmetric superradiant excitation of the array. When atoms are arranged in a lattice where the spacing is smaller than the wavelength of incident light, collective subradiant excitations begin to play an important role~\cite{vladimirova1998ru,Poddubny2008prb, Poddubny2009,Albrecht2019,kornovan2019extremely}.

The physics of such systems becomes especially rich in the multi-excitation regime due to photon blockade. Since a single qubit cannot be excited twice, the interaction between excitations becomes a decisive factor that strongly affects both the lifetime and spatial distribution of the collective many-body states. For arrays of two-level atoms,  subradiant two-excitation states are fermionized due to interactions~\cite{Molmer2019} and two-particle excitations which are products of dark and bright single-excitation states can appear~\cite{Ke2019arXiv}. Spatially bound subradiant dimers have also been predicted~\cite{Zhang2019arXiv} and a transition from few-body quantum to nonlinear classical regimes has recently been theorised~\cite{Mahmoodian2019}. 
\begin{figure}[!b]
\centering\includegraphics[width=0.48\textwidth]{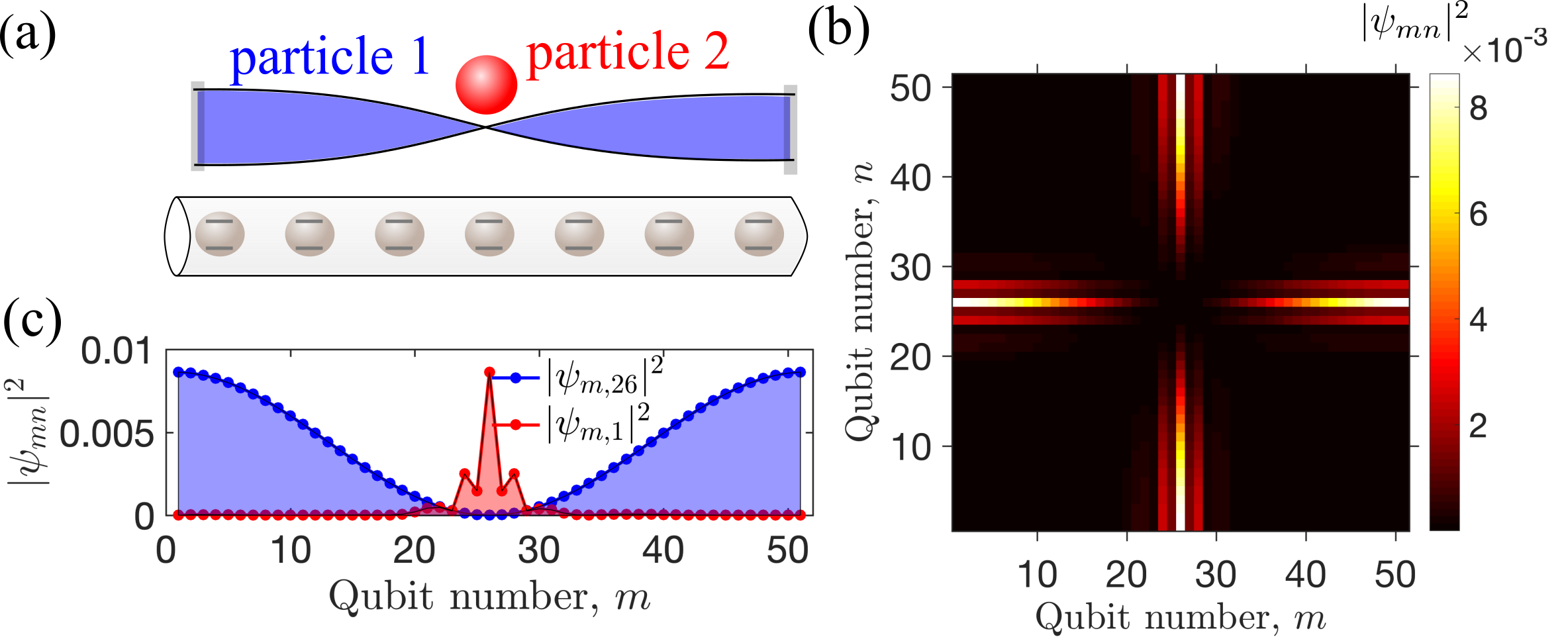}
\caption{(a) Schematic illustration of a two-particle state in an array of qubits in a waveguide, where one photon behaves as a standing wave and traps another one in the antinode of that wave. (b) Spatial map of  the corresponding two-excitation wavefunction $|\psi_{mn}|^2$ with $\eps/\Gamma_0=-2.57-0.54\rmi$ depending on the coordinates of the first and second qubits.  (c) Cross-sections of the map for $n=1$ and $26$. The calculation parameters are $\varphi=0.05$, $N=51$,  $\chi\to\infty$.}
\label{fig:scheme}
\end{figure}
%%%%%%%%%%%%%%%%%%%%%%%%%%%%%

In this Letter, we uncover and study a new class of two-particle hybrid excitations in arrays of subwavelength-spaced
two-level qubits coupled to a waveguide.  We reveal that when one of two indistinguishable  photons forms a standing
wave, the second photon can be localized in the nodes of this wave, as shown in Fig.~\ref{fig:scheme}(a). This effect can be viewed as a self-induced quantum optical lattice. This state is represented as a special type of photon-mediated cross-shaped states with strong spatial localization in the quasi-2D probability distribution, as in Figs.~\ref{fig:scheme}(b,c).
In the quasi-2D color map  Fig.~\ref{fig:scheme}(b) the $x$- and $y$-coordinates correspond to the positions of the first and second excitations, and the color  represents the probability of a pair to occupy that site. The ``cross shape''  means that the motion is highly constrained for the first excitation and free to propagate in space for the second excitation, or vice versa.  We demonstrate that such cross-shaped states arise naturally for subwavelength arrays in a broad range of
parameters which should be experimentally observable in systems where the qubits are probed individually~\cite{Wang2019}.

In the single-excitation regime for the same system, all single-particle states are ordinary delocalised standing waves. Thus, the presence of strong localization on just a few lattice sites in the two-excitation regime, as observed in Fig.~\ref{fig:scheme}(b), is solely a result of the interaction due to photon blockade. While the most subradiant excitations in considered system behave  as hard-core bosons~\cite{Molmer2019}, the ansatz of Ref.~\cite{Molmer2019} involves only  single-particle states that are delocalized and do not describe our cross-shaped states.  Since the localization involves two indistinguishable entangled particles, it is also qualitatively different from the physics of self-localized polarons which originates from electron-phonon interaction in solids. 

The spatially localized structure studied in this Letter bears some resemblance to the profiles of intrinsic localized modes known to occur in discrete systems~\cite{Campbell2004}, self-trapped localized solitons studied in nonlinear discrete systems~\cite{Kivshar1993}, compact localized states  in tight-binding models with flat bands~\cite{leykam2018artificial} as well as localized excitations found in generalized Bose-Hubbard models~\cite{Longhi2013,Gorlach2017}. However, in our system both long-range coupling and photon blockade are crucial for localization which distinguishes it from these studies.

%%%%%%%%%%%%%%%%%%%%%%%%%%%%%%%%%%%%%
{\it Model and numerical results.}
We consider $N$ periodically spaced qubits in a one-dimensional waveguide. In the Markovian approximation, this system is characterized by the Hamiltonian
 \begin{align}\label{eq:HM}
 \mathcal H=\sum\limits_{m,n}H_{m,n}b_{m}^{\dag}b_{n}+\frac{\chi}{2}\sum\limits_{n}b_{n}^{\dag} b_{n}^{\dag}b_{n}^{\vphantom{\dag}}b_n^{\vphantom{\dag}}\:,
 \end{align}
 where
\begin{equation}\label{eq:H}
H_{mn}\equiv-\rmi\Gamma_0\e^{\rmi \varphi |m-n|}\:,\quad m,n=1\ldots N\:.
\end{equation}
Here, $b_{m}$ are the  annihilation operators for the bosonic  excitations of the qubits and the parameter $\chi$ describes the interaction. The results for two-level qubits can be obtained in the limit of $\chi/\Gamma_0\to\infty$. The phase $\varphi=q_0d$ is given by the product of the  distance between the qubits $d$ and the light wave vector  $q_0$ at the qubit frequency. The parameter $\Gamma_0$ is the radiative decay rate of an individual qubit. We are interested in the spatial distribution of the two-excitation states of this system $\sum\psi_{mn} b_n^\dag b_m^\dag |0\rangle$ in the strongly subwavelength regime with $0<\varphi\ll 1$.
In the limit of $\chi\to\infty$, when $\psi_{nn}\equiv 0$, the  Schr\"odinger equation  can be presented in the following matrix form:
\begin{equation}\label{eq:S2}
H\psi+\psi H-2\:{\rm diag\:}[{\rm diag\:} H\psi]=2\eps \psi\:,
\end{equation}
with    $\psi_{nm}=\psi_{mn}$. Here, the first two terms in the left-hand side describe the propagation of  the first and second particle, respectively, and the third term accounts for the interaction.

\begin{figure}[t]
\centering\includegraphics[width=0.48\textwidth]{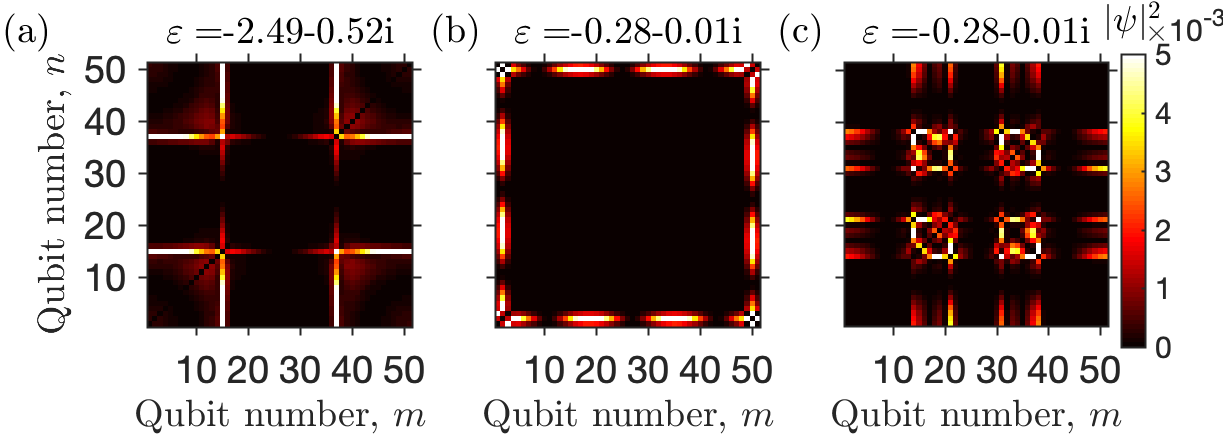}
 \centering \includegraphics[width=0.48\textwidth]{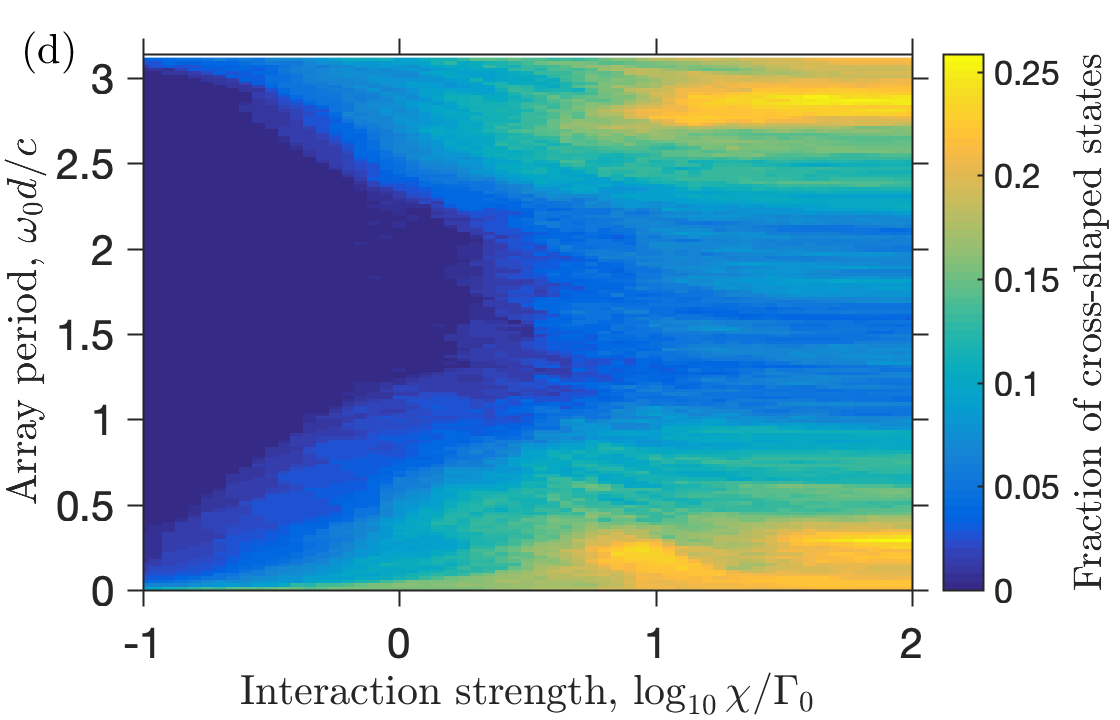}
\caption{(a-c): Examples of spatial distributions for different two-excitation states calculated for $N=51$ and $\varphi=0.01$.
Complex eigenenergy is shown for each panel. (d) Phase diagram showing the relative number of  cross-shaped  states Eq.~\eqref{eq:svd} depending on the interqubit phase $\varphi=\omega_0d/c$ and the interaction strength $\chi$. Calculated for $N=25$. }
\label{fig:numerics}
\end{figure}

Our numerical calculation demonstrates that a large number of two-excitation  states of the system Eq.~\eqref{eq:S2} have the following structure:
\begin{equation}
\psi_{nm}\approx \psi^{\rm (loc)}_n\psi^{\rm (free)}_m+\psi^{\rm (loc)}_m\psi^{\rm (free)}_n\:.\label{eq:svd}
\end{equation}
Here, the vector $\psi^{\rm (free)}_m$ is essentially a standing wave where the wave vector $\sim (1\div 4)\pi/N$ is slightly modified by the interaction. The vector $\psi^{\rm (loc)}_n$ has a very different shape and consists of peaks localized at just several sites which are pinned to the antinodes of the standing wave $\psi^{\rm (free)}_m$. Examples of several such states are presented in Fig.~\ref{fig:numerics}(a-c).
 Figure~\ref{fig:numerics}(d) shows how the number of cross-shaped states of the type Eq.~\eqref{eq:svd} depends on the distance between the qubits and the interaction strength. The states were singled out by requiring the inverse participation ratios for $\psi^{\rm (loc)}$ in real space and $\psi^{\rm (free)}$ in reciprocal space to be larger than 0.12, and simultaneously requiring the remaining singular values in the Schmidt decomposition to be smaller than 0.25. The cross-like states occupy up to 25\% of the two-excitation spectrum when the phase is close to an integer multiple of $ \pi$ and when the photon-photon interaction is strong such that $\chi\gg \Gamma_0$.

%%%%%%%%%%%%%%%%%%%%%%%%%%%%%
%%%%%%%%%%%%%%%%%%%%%%%%%%%%%

\begin{figure}[!t]
\centering\includegraphics[width=0.48\textwidth]{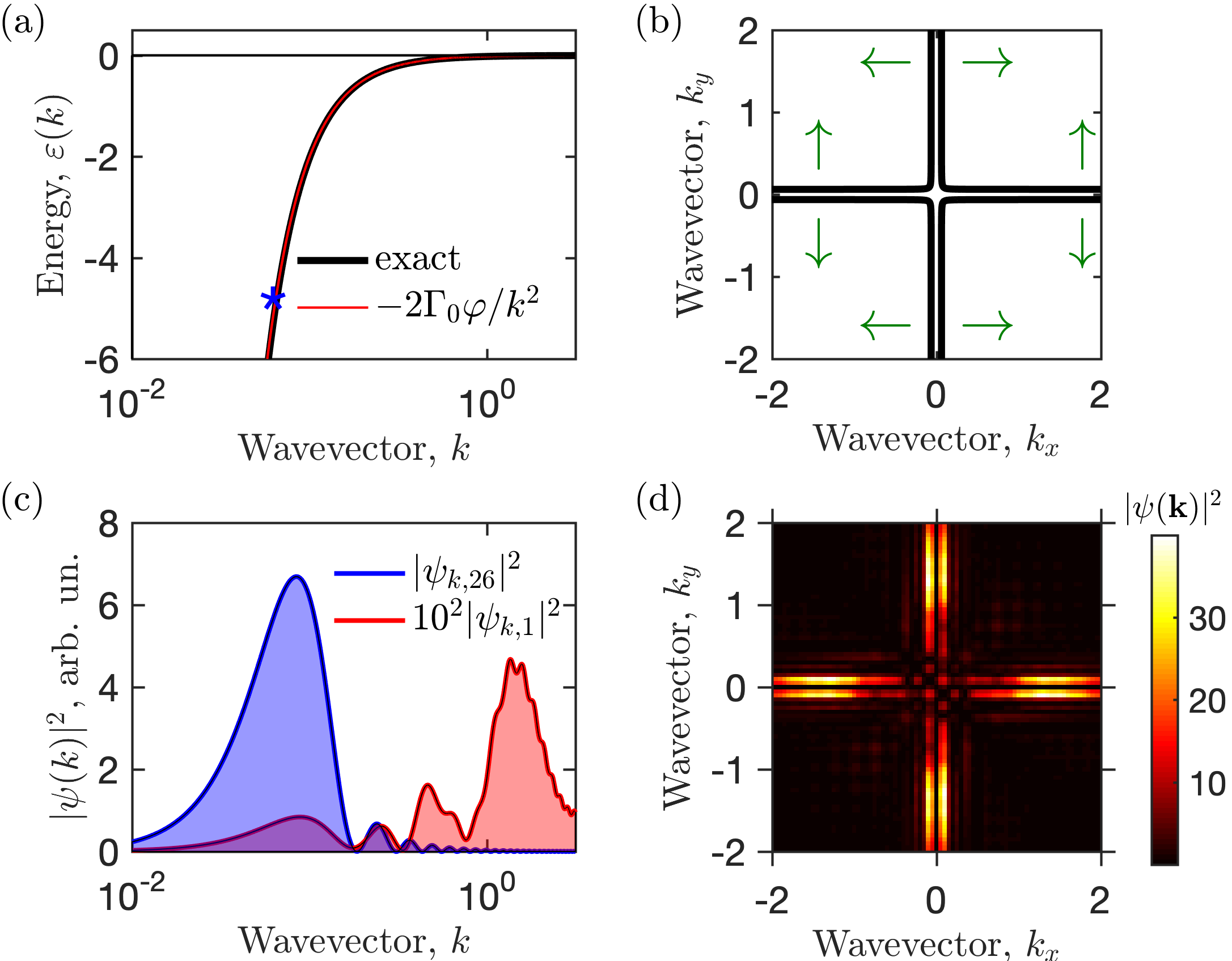}
\caption{(a) Lower polariton branch of the single-particle dispersion. Star indicates the energy $\eps=-2\times 2.57\Gamma_0$ and the wave vector corresponding to the cross-shaped state in Fig.~\ref{fig:scheme}.
 (b) Two-excitation dispersion Eq.~\eqref{eq:iso} for $\eps=-2.57\Gamma_0$. Green arrows indicate the directions of group velocity in the corresponding regions.
(c,d) One- and two-dimensional Fourier transforms of the state $\psi_{mn}$ in Fig.~\ref{fig:scheme}.
Blue and red curves in (c) are calculated for the $m$ index  being at the center or at the edge, respectively.
 Calculation parameters are the same as in Fig.~\ref{fig:scheme}.}
\label{fig:kxky}
\end{figure}

%%%%%%%%%%%%%%%%%%%%%%%%%%%%%%%%%%%%%
{\it Quasi-flat polaritonic dispersion.}
We focus on the cross-like state shown in Fig.~\ref{fig:scheme}.  It is instructive to first review the results for single-particle states in an infinite periodic array $\sum_n\e^{\rmi kn} \sigma_n^\dag |0\rangle$, where $k$ is  the Bloch wave vector.
These states are coupled light-matter excitations (polaritons) and their energy dispersion is given by $ \eps(k)=\Gamma_0\sin \varphi/(\cos k-\cos\varphi)$~\cite{Ivchenko1991}. The dispersion consists of upper and lower polariton branches separated by the gap around the qubit resonant frequency. We are interested only  in the states of the lower polaritonic branch in the regime where $
 \varphi\ll k\ll 1\:.
$
 In this case, the lower polaritonic branch can be well approximated by
  \begin{equation}\label{eq:epsk}
 \eps(k)\approx -\frac{2\varphi\Gamma_0}{k^2}\:,\quad k\ll 1\:,
 \end{equation}
 as demonstrated by the calculation in Fig.~\ref{fig:kxky}(a). The important feature of the dispersion Eq.~\eqref{eq:epsk} which is central for our study is that there is a huge density of states for $\eps$ just below zero when the group velocity is small and $\rmd k/\rmd \eps \gg 1$. In other words, the polaritons are heavy and slow, which strongly facilitates their trapping. This single-particle dispersion leads to quite interesting isoenergy contours  for a pair  of noninteracting polaritons with the given total energy $2\eps$ and the wave vectors $k_x,k_y$. The isoenergy contour is given by
  \begin{equation}
 \eps(k_x,k_y)\approx-\varphi \Gamma_0\left(\frac1{k_x^2}+\frac1{k_y^2}\right)\:\label{eq:iso}
 \end{equation}
 and is plotted in Fig.~\ref{fig:kxky}(b) for the average pair energy  $\eps=-2.57\Gamma_0$ corresponding to the real part of the complex energy of the state in Fig.~\ref{fig:scheme}(b). Crucially, for most points of the isofrequency contour, the group velocity $\rmd \eps/\rmd \bm k$ is parallel to either $x$- or $y$- axis. This means that only one of the two photons can propagate in space at the same time, in full agreement with the real space maps of the eigenstates shown in Figs.~\ref{fig:scheme},\ref{fig:numerics}.

It is instructive to study  the wavefunction shown in Fig.~\ref{fig:scheme} in the reciprocal space.  The results are presented in Fig.~\ref{fig:kxky}c and  Fig.~\ref{fig:kxky}d.  We start by calculating the Fourier transform along only one particle coordinate, $|\sum_n\e^{-\rmi kn}\psi_{mn}|^2$ when the second particle position $m$ is either at the center ($m=26$) or at the edge  ($m=1$). Indeed, the Fourier transform along the center
reveals a sharp peak that corresponds to a standing wave with a well-defined wave vector (blue curve in  Fig.~\ref{fig:kxky}c). The Fourier transform  at the edge results in a broad distribution of large wave vectors characteristic for a localized state (red curve in  Fig.~\ref{fig:kxky}c). The same results can be deduced from the two-dimensional Fourier transform $|\sum_{nm}\e^{-\rmi k_xm-\rmi k_yn}\psi_{mn}|^2$ plotted in  Fig.~\ref{fig:kxky}(d): one of the two polaritons has large wave vector when the other one has a small one, or vice versa.

Interestingly, the numerically obtained properties of the cross-like states seem to be in general agreement with the features of our study of metastable twilight states reported in Ref.~\cite{Ke2019arXiv}. In this paper, the twilight state is defined as a metastable product of dark and bright states. Here, the cross-like states result from the products of less subradiant states (standing wave) with  strongly subradiant states (localized distribution).  In this Letter, we focus  on the spatial distribution  of the two-photon states, rather than their lifetimes, but more details are given in the Supplementary Materials.

%%%%%%%%%%%%%%%%%%%%%%%%%%%%%
%%%%%%%%%%%%%%%%%%%%%%%%%%%%%

{\it Interaction-induced localization.} The flat dispersion is an essential ingredient for the trapping of polaritons.
The second necessary ingredient is their interaction. In order to explain the interaction effects analytically we have to overcome the technical difficulty of the original Schr\"odinger equation Eq.~\eqref{eq:S2} which has a very dense Hamiltonian for $\varphi\ll 1$, $H_{nm}\approx -\rmi \Gamma_0$ due to the long-range coupling between the qubits. It is useful to write a two-particle equation for the matrix $\mathcal E=H\psi H$ rather than the two-photon wavefunction $\psi$ directly.  By rewriting the Schr\"odinger equation under this tranformation (see Supplementary Materials for details), we can derive an equation of the form
 \begin{equation}
{(\partial_{x}^{2}+\partial_{y}^{2}){\mathcal E}-\delta_{x,y}( \partial_{x}^{2}+\partial_{y}^{2}){\mathcal E}=
 \frac{\eps}{\phi \Gamma_0}  \partial_{x}^{2} \partial_{y}^{2}{\mathcal E} \label{eq:chi}}\:,
 \end{equation}
 where $x,y=1,N$ and   $\partial_{x,y}^2$ are just the operators of discrete second-order derivatives,
 $\partial_x^2= \partial^2\otimes 1_{N\times N},\:\partial_y^2=  1_{N\times N}\otimes \partial^2$. The $N\times N$ matrix $\partial^2$ is defined as
 \begin{equation}
 \partial^{2}\equiv \frac{1}{2}\left(\begin{smallmatrix}
 -1&1&0&\ldots\\
 1&-2&1&\ldots\\
  &&\ddots&\\
  \ldots&1&-2&1\\
 \ldots  &0& 1&-1
 \end{smallmatrix}\right)\:.
 \end{equation}
 Importantly, in Eq.~\eqref{eq:chi} we have neglected the radiative decay of the eigenstates, $\Im \eps\equiv 0$, which is a reasonable approximation in the considered strongly subwavelength regime with $\varphi\ll 1$.  If the interaction term $\propto \delta_{x,y}$ is omitted, the eigenstates of Eq.~\eqref{eq:chi} are just  standing waves with the dispersion law Eq.~\eqref{eq:iso}. We have verified numerically that when the interaction term is kept, Eq.~\eqref{eq:chi} features the same kind of cross-shaped eigenstates as our original Schr\"odinger equation  Eq.~\eqref{eq:S2}. Hence, while Eq.~\eqref{eq:chi} looks quite simple, it still captures the physics of the interaction-induced localization. Importantly, since the matrices $\partial_{x,y}^2$ are tri-diagonal, Eq.~\eqref{eq:chi} is local in both the first and second photon coordinates $x$ and $y$. The physical reason why Eq.~\eqref{eq:chi} is local is that the matrix $\mathcal E$ describes a two-photon amplitude of the electric field in contrast with the matrix of two-qubit excitations $\psi_{mn}$. The considered array of qubits is subwavelength and can be viewed as a quantum nonlinear metamaterial~\cite{Ustinov2014}. As such, it is natural to expect a local   two-photon wave equation in the effective-medium approximation.
   %%%%%%%%%%%%%%%%%%%%%%%%%%%
    \begin{figure}[t]
\centering \includegraphics[width=0.48\textwidth]{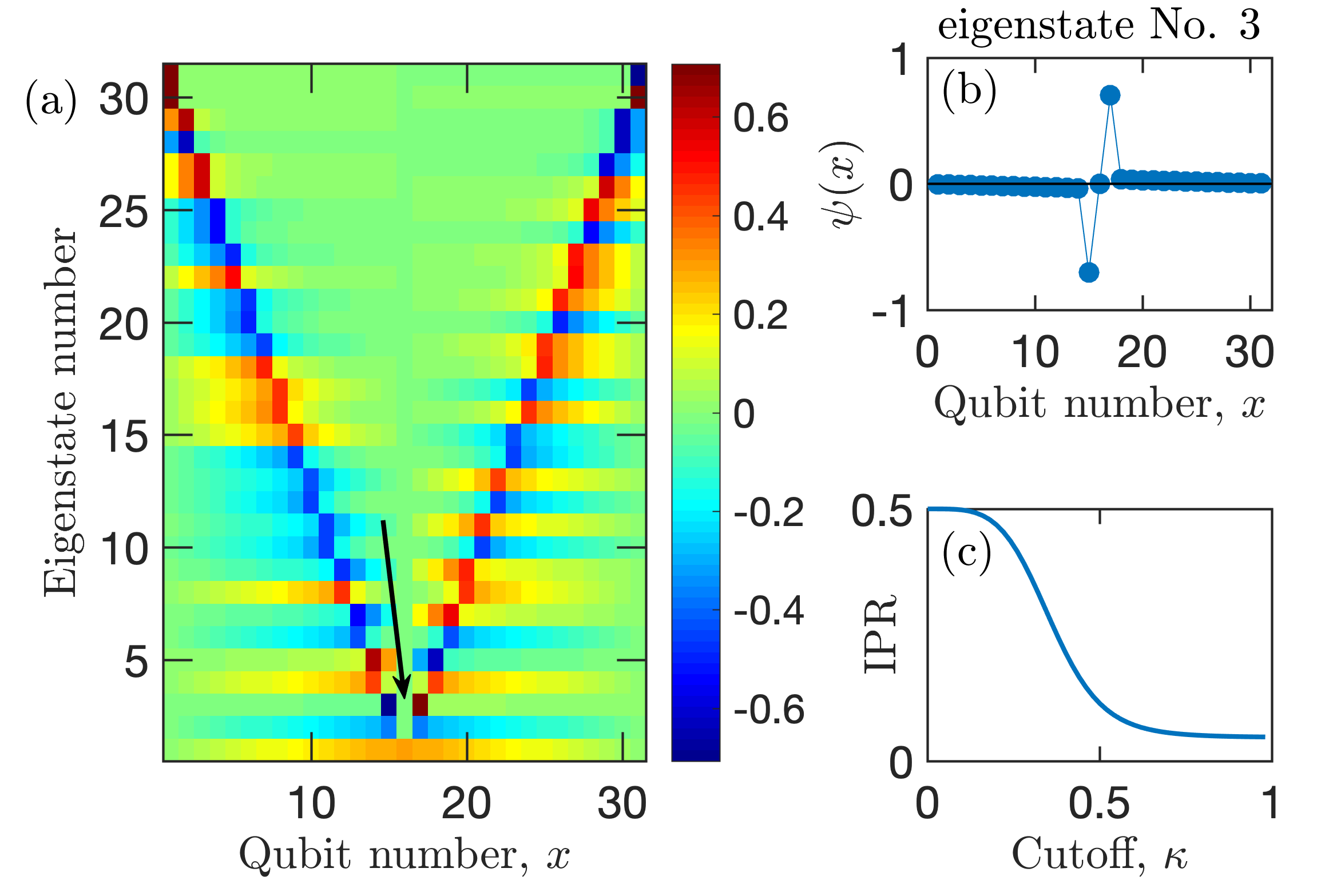}
 \caption{(a) Eigenstates of the operator \eqref{eq:L} for $\kappa=0.1$, $N=31$.
 (b) Spatial profile of the third eigenstate, indicated in (a) by an arrow. Inverse participation ratio (IPR) of the third eigenstate depending on the cutoff parameter $\kappa$.}\label{fig:L}
 \end{figure}
 %%%%%%%%%%%%%%%%%%%%%%%%%%%

 In order to explain the localization it remains only  to understand why the diagonal cross section of the two-photon  distribution $\mathcal E(x,x)$ is localized at $x\approx N/2$. To this end, we introduce the Green's function
$G(x,y;x',y';\eps)$ of Eq.~\eqref{eq:chi} without the interaction term, satisfying
   \begin{equation}
 (\partial_{x}^{2}+\partial_{y}^{2})G=
 \frac{\eps}{\varphi\Gamma_0} \partial_{x}^{2} \partial_{y}^{2}G+\delta_{x,x'}\delta_{y,y'}\:.\label{eq:defG}
 \end{equation}
 The solution of Eq.~\eqref{eq:defG} can be expanded over the single-particle eigensolutions which are just standing waves. We are now interested in the case when the photon pair energy $2\eps$ is close to the resonance of the given standing wave $u_0$ with the wave vector $k_0$ and the energy $\eps_0\approx 2
\eps$.
 The Green's function can then be approximated by the following  general expression
  \begin{multline}
 G(x,y;x',y';\eps)\approx \frac{a}{\delta\eps}\\\times
[ u_0(x)u_0(x') g(y,y')+ u_0(y)u_0(y') g(x,x')],\label{eq:gres0}
  \end{multline}
  where $\delta\eps=\eps-\eps_0/2$ and $a=\Gamma_0/k_0^2$.  Here, the matrix
  $g(y,y')$ describes the short-range components of the Green's function.  By construction, the distribution
$ G$ as a function of $x,y$ for given $x',y'$  has a cross-like shape which is characteristic of the isoenergy contours discussed in Fig.~\ref{fig:kxky}.
   More detailed derivation and analysis of Eq.~\eqref{eq:gres0} is presented in the Supplementary Materials,  where we demonstrate that the short-range component can be qualitatively approximated by
  $g(y,y')\approx  [\partial^{2}-\kappa^2 ]^{-1}_{y,y'}$, where $\kappa \sim k_0\ll 1$ is a cutoff parameter.
  Physically, the function $g(y,y')$ takes into account the net contribution to the Green's function from all standing waves with the wave vectors larger than $k_0$.

  Substituting Eq.~\eqref{eq:gres0} into Eq.~\eqref{eq:chi} we obtain the following equation for the diagonal components of the matrix $\mathcal E$:
\begin{equation}
\delta\eps\mathcal E(x,x)=\mathcal L_{x,x'}\mathcal E(x',x')\:,\label{eq:E}
 \end{equation}
where
   \begin{equation}
\mathcal L= 2a\diag[u _0(x)] [\partial_x^{2}-\kappa^2 ]^{-1}\diag[u _0(x)]\partial_{x}^{2}\label{eq:L}.
 \end{equation}
 It can be easily checked numerically that for $\kappa\ll 1$ the operator $\mathcal L$ has spatially localized eigenstates, see Fig.~\ref{fig:L}(a).
   Specifically, the third eigenstate, shown in Fig.~\ref{fig:L}(b),
 describes  the diagonal cross-section of the considered  cross-like distribution.
 Fig.~\ref{fig:L}(c) shows the inverse participation ratio (IPR) $\sum_x|\psi_x|^4$ for the third eigenstate as a function of the cutoff parameter $\kappa$.
 For $\kappa \ll 1$ the eigenstate is practically independent of the cutoff parameter and looks  like a  derivative of a discrete $\delta$-function.  One  can then interpret Eq.~\eqref{eq:E} as describing a motion of a particle with large mass in the potential  determined by $u _{0}(x)^{2}\propto \cos^2(k_0x)$. Clearly, in such a case the localization takes place in the node of the standing wave $u _0(x)$, in agreement with the results in Fig.~\ref{fig:L}. More detailed analysis is given in the Supplementary Materials.

In summary,  we have revealed that the presence of a polariton quantized as standing wave in a finite qubit array creates
an effective potential to trap the second polariton. This second polariton is pushed by the repulsive interaction to become localized in the antinodes of the standing wave, and stays trapped there due to its large effective mass. Our finding demonstrates that the interaction yields surprising results in the strongly quantum regime when only several particles are present in the system. We believe that the potential of waveguide quantum electrodynamical platforms for analog quantum simulations of many-body effects is still largely unexplored. For instance, it is not clear whether the considered states can be generalised to the many-body case, such as whether two photons can form an effective two-dimensional optical lattice that can trap the third photon in its nodes. Another open but very interesting  question is the role of interactions in topologically nontrivial qubit arrays~\cite{Poshakinskiy2014,Ozawa2019}.

%%%%%%%%%%%%%%%%%%%%%%%%%%%%%%%%%%%%%%%%%%%%%%%%%

%%%%%%%%%%%%%%%%%%%%%%%%%%%%%%%%%%%%%%%%%%%%%%%%%

\let\oldaddcontentsline\addcontentsline% Store \addcontentsline
\renewcommand{\addcontentsline}[3]{}% Make \addcontentsline a no-op

\begin{acknowledgments}
We acknowledge useful discussions with J. Brehm, M.M.~Glazov, L.E.~Golub, I.V.~Iorsh,  E.L.~Ivchenko, A.A.~Sukhorukov and A.V.~Ustinov. This work was supported by the Australian Research Council. C. Lee was supported by the National Natural Science Foundation of China (NNSFC) (grants
11874434 and 11574405). Y.~Ke was partially supported by the International Postdoctoral Exchange Fellowship Program (grant 20180052). A.V.P. also acknowledges a partial support from the Russian President Grant No. MK-599.2019.2. AVP, ANP and NAO have been partially supported by  the Foundation for the Advancement of Theoretical Physics and Mathematics ``BASIS''.
\end{acknowledgments}

%%%%%%%%%%%%%%
%\nocite{apsrev41Control}
%\bibliographystyle{apsrev4}
%\bibliography{titleon,cross}

\let\addcontentsline\oldaddcontentsline% Restore \addcontentsline

\newpage

\setcounter{figure}{0}
\setcounter{section}{0}
\setcounter{equation}{0}
\renewcommand{\thefigure}{S\arabic{figure}}
\renewcommand{\thesection}{S\Roman{section}}
\renewcommand{\thesection}{S\arabic{section}}
\renewcommand{\theequation}{S\arabic{equation}}
\renewcommand*{\citenumfont}[1]{S#1}
\renewcommand*{\bibnumfmt}[1]{[S#1]}

\begin{center}
\textbf{\Large Online supplementary information}
\end{center}
\tableofcontents
%%%%%%%%%%%%%%%%%%%%%%%
\section{Two-particle Schr\"odinger equation}
%%%%%%%%%%%
We start with the Schr\"odinger equation for the two-photon wavefunction
\begin{gather}\label{eq:S1}
H_{mn'}\psi_{n'n}+\psi_{mn'}H_{n'n}+\chi \delta_{mn}\psi_{nn}=2\eps \psi_{mn},\text{ or }\\
H\psi+\psi H+\chi {\rm diag\:}[{\rm diag\:} \psi]=2\eps \psi\:.
\end{gather}
We would now like to go to the limit $\chi\to\infty$ to get rid of $\chi$.
Importantly, even though $\psi_{nn}\to 0$ for $\chi\to\infty$, we still have
 $\chi \psi_{nn}\to\rm const$.
 The value of  $\chi \psi_{nn}$ can be calculated as a perturbation, see also Ref.~\cite{Ke2019arXiv}:
\begin{multline}
\chi \psi_{nn}=-\sum\limits_{n'\ne n}(H_{nn'}\psi_{n'n}+\psi_{nn'}H_{n'n})=\\-
2\sum\limits_{n'\ne n}H_{nn'}\psi_{n'n}=-2\sum\limits_{n'=1}^{N}H_{nn'}\psi_{n'n}\:.
\end{multline}
Hence, we can rewrite the Schr\"odinger equation in the limit $\chi\to \infty$ as
\begin{gather}
H_{mn'}\psi_{n'n}+\psi_{mn'}H_{n'n}-2\delta_{mn}H_{nn'}\psi_{n'n}=2\eps \psi_{mn},\\
H\psi+\psi H-2\:{\rm diag\:}[{\rm diag\:} H\psi]=2\eps \psi\:, \text{ with }\psi_{nn}=0\:,
\end{gather}
or,
in a matrix form,
\begin{equation}
H\psi+\psi H-2\:{\rm diag\:}[{\rm diag\:} H\psi]=2\eps \psi\:,\label{eq:Sh2}
\end{equation}
which is identical to Eq.~\eqref{eq:S2} in the main text.
In order to obtain Eq.~\eqref{eq:chi} in the main text we notice that for $\varphi\ll 1$
\begin{equation}
H^{-1}\approx \frac1{2\varphi}\partial^2,\text{ where } \partial^{2}\equiv \frac{1}{2}\left(\begin{smallmatrix}
 -1&1&0&\ldots\\
 1&-2&1&\ldots\\
  &&\ddots&\\
  \ldots&1&-2&1\\
 \ldots  &0& 1&-1
 \end{smallmatrix}\right)\:.\label{eq:ih}
\end{equation}
The physical picture behind Eq.~\eqref{eq:ih} can be understood by noticing that it is equivalent to the single-particle dispersion law 
  \begin{equation}\label{eq:epsk2}
 \eps(k)\approx -\frac{2\varphi\Gamma_0}{k^2}\:,
 \end{equation}
 in the infinite structure.  This dispersion law could be obtained  from  the exact dispersion $ \eps(k)=\Gamma_0\sin \varphi/(\cos k-\cos\varphi)$ in the limit $\varphi\ll k\ll 1$~\cite{Ivchenko1991}. Alternatively, it can be understood from the effective medium description of the qubit array, when
 it is described by the resonant permittivity
  \begin{equation} 
  \epsilon(\omega)=1+\frac{\Gamma_0}{2\varphi (\omega_0-\omega)}\:.
  \end{equation}
  Applying the wave equation $q^2=(\omega/c)^2\epsilon(\omega)\approx (\omega_0/c)^2\epsilon(\omega)$ in the vicinity of the frequency $\omega_0$, we obtain 
  $\omega-\omega_0=-2\varphi\Gamma_0/(qd)^2\:,$ in agreement with Eq.~\eqref{eq:epsk2}.

Substituting $\psi=H^{-1}\mathcal EH^{-1}$ into Eq.~\eqref{eq:Sh2} we find 
\begin{equation}
H^{-1}\mathcal E+\mathcal EH^{-1}-2\:{\rm diag\:}[{\rm diag\:} \mathcal E H^{-1}]=2\eps H^{-1}\mathcal EH^{-1}\:,\label{eq:Sh3}
\end{equation}
which, given Eq.~\eqref{eq:ih},
reduces to Eq.~\eqref{eq:chi} in the main text.

        %%%%%%%%%%%%%%%%%%%%%%%%%%%
    \begin{figure*}[t!]
\centering \includegraphics[width=0.95\textwidth]{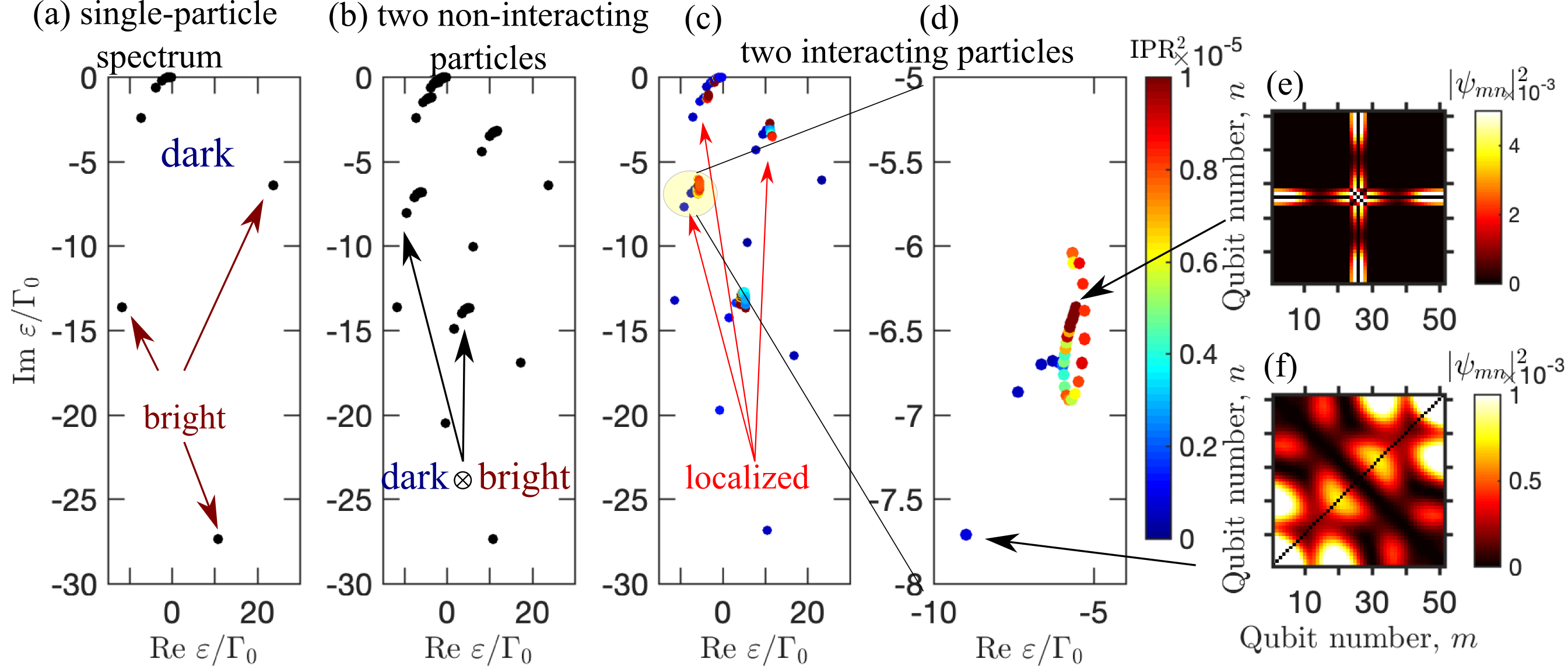}
 \caption{Complex energy spectrum of array with $N=51$ qubits and $\phi=\omega_0d/c=0.4$. (a) single-particle spectrum. (b) spectrum of non-interacting pairs of particles. (c) spectrum for interacting pairs, $\chi/\Gamma_0\to\infty$; (d) same as (c), but zoomed in. Panels (c) and (d) are colored by the square of inverse participation ratio, red means stronger localization.  (e,f) false color map of the wavefunction for the most- and the least- localized states from (d), indicated by arrows. }\label{fig:lifetimes}
 \end{figure*}
    %%%%%%%%%%%%%%%%%%%%%%%%%%%
\section{Complex energy spectrum of the cross-like states}
    %%%%%%%%%%%%%%%%%%%%%%%%%%%
    In order  to elucidate the considered interaction-induced  localization, it is instructive to analyze the complex energy spectrum of the qubit array. Figure~\ref{fig:lifetimes}  presents single- and double-excitation energy spectrum for the array with $N=51$ qubits and $\phi=\omega_0d/c=0.4$. Panel (a) shows the single-particle spectrum. It consists of several bright modes (indicated by arrows) and a large number of subradiant states. All these subradiant states are delocalized and correspond just to standing waves of the lower polaritonic branch. Now, in Fig.~\ref{fig:lifetimes}(b) we present the mean energies for non-interacting pairs of particles. These can be obtained by either setting $\chi=0$ in the two-photon Hamiltonian Eq.~\eqref{eq:S1} or just taking sums for all possible pairs of single particles complex energies, $\eps=(\eps_{n}+\eps_{m})/2$.  The spectrum in Fig.~\ref{fig:lifetimes}(b) has quite a special structure: it consists of several distinct clusters of close-lying eigenenergies. These clusters originate from a given single-particle bright mode from Fig.~\ref{fig:lifetimes}(a) combined with the single-particle dark modes. Hence, the shape of each cluster repeats the shape of the cluster of single-particle dark modes in Fig.~\ref{fig:lifetimes}(a). 
    In Fig.~\ref{fig:lifetimes}(c) we  examine, how this complex spectrum of pairs in  Fig.~\ref{fig:lifetimes}(b) is affected by the interactions, $\chi/\Gamma_0\to\infty$. The calculation demonstrates, that the overall shape of the spectrum in the presence of interaction remains the same. 
    It addition to the true subradiant states with $|\Im\eps|\lesssim \Gamma_0$, the spectrum consists of distinct clusters. Modes in  each cluster stem from products of dark and bright states and can be identified as {\it twilight} modes, following our previous work \cite{Ke2019arXiv}. However, in contrast to Ref.~\cite{Ke2019arXiv}, where we have not looked into the spatial structure of the eigensolutions, we now focus on their spatial localization degree. Namely, each point in Fig.~\ref{fig:lifetimes}(c) is now colored according to the inverse participation ration of the corresponding eigenstate, $\sum_{m,n}|\psi_{m,n}|^4$. 
    It is now clearly seen that each cluster has several red-coloured states with  stronger localization degree. In  Fig.~\ref{fig:lifetimes}(d) we show the  spectrum for a given single  cluster of twilight states in a larger scale. The spectrum has quite a complex shape, that qualitatively differs from the spectrum of non-interacting particles in     Fig.~\ref{fig:lifetimes}(a,b). As such, these states can not be described by the fermionic ansatz of Refs.~\cite{Molmer2019,Zhang2019arXiv}. Fig.~\ref{fig:lifetimes}(e) shows the wavefunction for the most localized state from  Fig.~\ref{fig:lifetimes}(d), indicated by an arrow. This state has a characteristic cross-like shape. It is similar to Fig.~\ref{fig:scheme}(b) in the main text and very different from that for the conventional twilight state, shown in Fig.~\ref{fig:scheme}(f).         We conclude, that the interactions are drastically important for some of the twilight eigenstates in a finite array of qubits and lead to qualitative modification of their spatial structure.
    
%%%%%%%%%%%%%%%%%%%%%%%
\section{Green's function for non-interacting particles}
%%%%%%%%%%%%%%%%%%%%%%%
    %%%%%%%%%%%%%%%%%%%%%%%%%%%
    \begin{figure}[t]
\centering \includegraphics[width=0.45\textwidth]{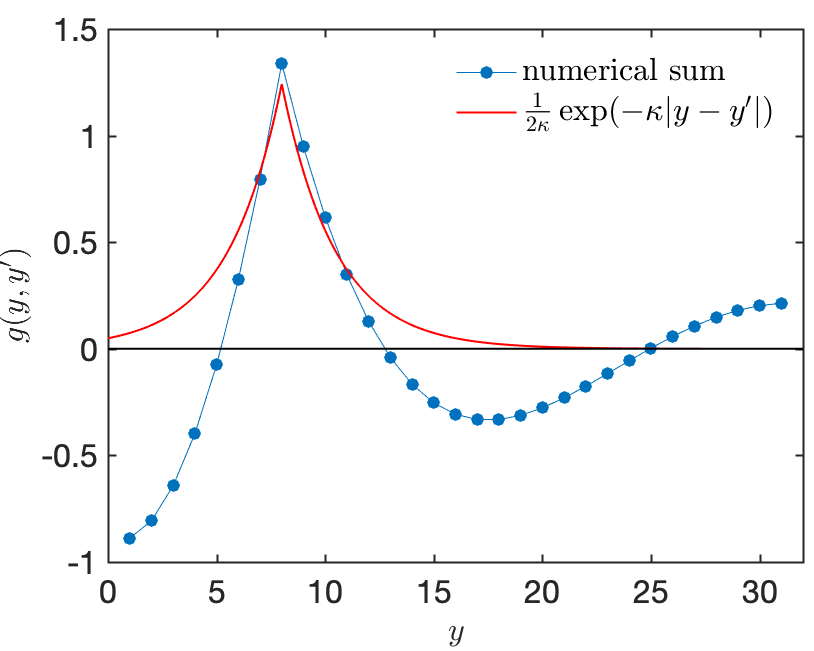}
 \caption{Short-range component of the Green's function calculated numerically  from Eq.~\eqref{eq:sumg} and analytically from Eq.~\eqref{eq:g1}. The calculation has been performed for $N=31$ qubits, $k_{n_0}=3
 \pi/N$, $\kappa=4\pi/N$ and $y'=8$.}\label{fig:g}
 \end{figure}
In this section we analyze the Green's function of  Eq.~\eqref{eq:defG} of the main text. Our goal is to understand, why it favours  the cross-like spatial patterns and can be approximated by Eq.~\eqref{eq:gres0} from the main text.  The equation for the Green's function is reiterated below
   \begin{multline}
 (\partial_{x}^{2}+\partial_{y}^{2})G(x,y;x',y';\eps)\\=
 \frac{\eps}{\varphi\Gamma_0} \partial_{x}^{2} \partial_{y}^{2}G(x,y;x',y';\eps)+\delta_{x,x'}\delta_{y,y'}\:.\label{eq:defG1}
 \end{multline}
 The solution can be approximated by a sum of the eigenstates of the single-particle equation
 \begin{equation}
  \partial_{x}^{2}u_{n}(x)=-k_{n}^{2}u_{n}(x)
 \end{equation}
in the  following way
    \begin{equation}\label{eq:sumG}
 G(x,y;x',y';\varepsilon)\approx-\sum\limits_{n,m=1}^{N}\frac{u_{n}(x)u_{n}(x')u_{m}(y)u_{m}(y')}{(\eps/\varphi\Gamma_0)k_{n}^{2}k_{m}^{2}+k_{n}^{2}+k_{m}^{2}}\:.
 \end{equation}
 We skip the terms with $n,m=0$ from the sum, since they have  $ \partial_{x}^{2}u_0\equiv 0$ and cancel out when the expansion Eq.~\eqref{eq:sumG} is substituted into Eq.~\eqref{eq:defG1}.
 The single-particle  solutions are  approximately given by
 \begin{gather}
 k_{n}\approx \frac{\pi}{N}(n-1),\quad n=1\ldots N,\\\nonumber
 \quad u_{n}(x)= c_{n}\cos k_{n}(x-1/2),\quad c_{n}=\sqrt{\frac{1+\delta_{n,1}}{N}}\:.
 \end{gather}
 and are just the standing waves. Importantly, we consider the states with  the energy close to the resonance of the given standing wave,
\begin{equation}
 \eps_0= -\frac{2\Gamma_0}{\varphi k_{n_{0}}^{2}}, \quad n_0\sim 1\ll N.\label{eq:resonance}
 \end{equation}
 For example, the state in Fig.~\ref{fig:scheme}(b) in the main text corresponds to  $n_0=2$, the corresponding single-particle solution is just $\cos (\pi n/N)$.
Using Eq.~\eqref{eq:resonance} and the fact that the energy of a pair of particles 
  \begin{equation}
 \eps(k_x,k_y)\approx-\varphi \Gamma_0\left(\frac1{k_x^2}+\frac1{k_y^2}\right)\:\label{eq:iso1}
 \end{equation}
does not depend on $k_{y}$ for $k_{y}\gg k_{x}\equiv k_{n_0}$, we write
  \begin{multline}
 G(x,y;x',y';\eps)=-\sum\limits_{n,m=1}^{N}(\ldots)\approx
 \sum\limits_{n=n_0,m\gg n_0}(\ldots)\\+ \sum\limits_{m=n_0,n\gg n_0}(\ldots)   \end{multline}
 As a result, the series Eq.~\eqref{eq:sumG} are simplified to
 \begin{multline}
 G(x,y;x',y';\eps)\approx \frac{1}{k_{n_{0}}^{2}(\eps-\eps_0/2)+1}\\\times
[ u_{n_{0}}(x)u_{n_{0}}(x') g(y,y')+ u_{n_{0}}(y)u_{n_{0}}(y') g(x,x')],\label{eq:gres}
  \end{multline}
  where
   \begin{equation}
\quad g(y,y')=\sum\limits_{m=n_{\rm min}}^{N}
 \frac{u_{m}(y)u_{m}(y')}{k_{m}^{2}} \label{eq:sumg}
  \end{equation}
  and $n_{\rm min}\gtrsim n_{0}$.
  This is equivalent to Eq.~\eqref{eq:gres0} in the main text.
Our numerical calculation, presented at   Fig.~\ref{fig:g}, demonstrates that  at short distances
  \begin{equation}\label{eq:g1}
  g(y,y')\approx \frac1{2\kappa}\e^{-\kappa |y-y'|},\kappa\approx k_{n_{\rm min}+1}\:,
  \end{equation}
  In another words, the operator $g$ looks like
  \begin{equation}\label{eq:g2}
  g\propto (\partial_x^2-\kappa^2)^{-1}
  \end{equation}
  with $k_{n_0}\lesssim \kappa \ll 1$.
  Physically, this result can be interpreted as follows. Let us consider the terms with $m\ll N$, when the summation in Eq.~\eqref{eq:sumg} can be approximated  by the integration. Since the functions $u_{m}(y)$ are just plane waves, and $k_m\propto m$, we would have $g\propto \int \rmd k \cos (ky) \cos (ky')/k^2$. This is just  the double-integrated $\delta$-function of $y-y'$, so $g(y,y')\propto |y-y'|$. On the other hand, at larger distances there exists a cutoff because the series starts from some non-zero wave-vector $\sim\kappa$. This cutoff leads to the replacement of the function $|y-y'|$ by a decaying exponent $\e^{-\kappa |y-y'|}$, and results in the Green's function of type Eq.~\eqref{eq:g1}.

  %%%%%%%%%%%%%%%%%%%%%%%%%%%
    \begin{figure}[t]
\centering \includegraphics[width=0.5\textwidth]{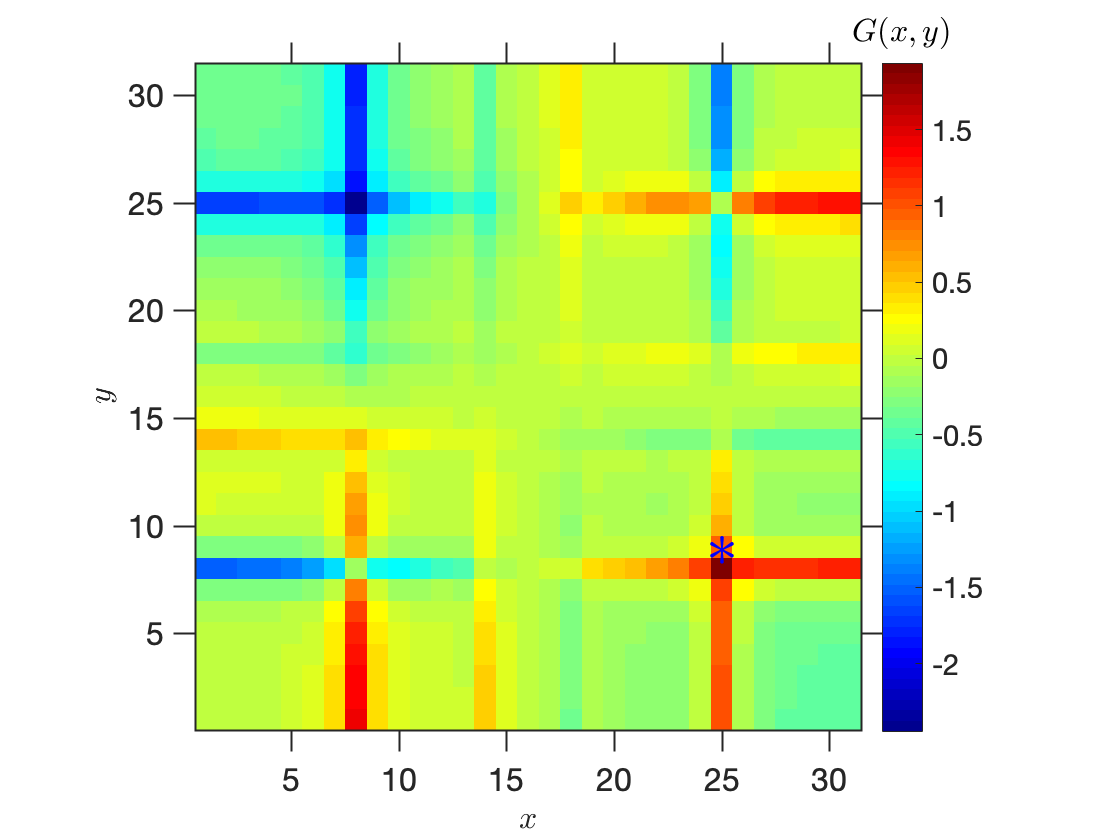}
 \caption{Numerically calculated Green's function of Eq.~\eqref{eq:defG1}. Calculation has been performed for the array with $N=31$ qubits. The source has been located at the point $x'=25,y'=8$, indicated by a start, and the energy is $\eps=-194\varphi\Gamma_0$.}\label{fig:G}
 \end{figure}

   %%%%%%%%%%%%%%%%%%%%%%%%%%%
   It can be verified numerically or analytically, that $G(x,y;x',y';\eps)$ as a function of $x$ and $y$  for fixed
   $x',y'$ has a cross-like feature, see Fig.~\ref{fig:G}. In another words, localized defects induce cross-like distributions even without the interaction. The presence of at least four crosses in the distribution in Fig.~\ref{fig:G} instead of one could be another finite size effect. It could be caused by  the reflection of the waves from the edges of the structure and requires further analysis.

   %%%%%%%%%%%%%%%%%%%%%%%%%%%
   \section{Approximate solution for localized eigenstates}
  %%%%%%%%%%%%%%%%%%%%%%%%%%%
      \begin{figure}[t]
\centering \includegraphics[width=0.4\textwidth]{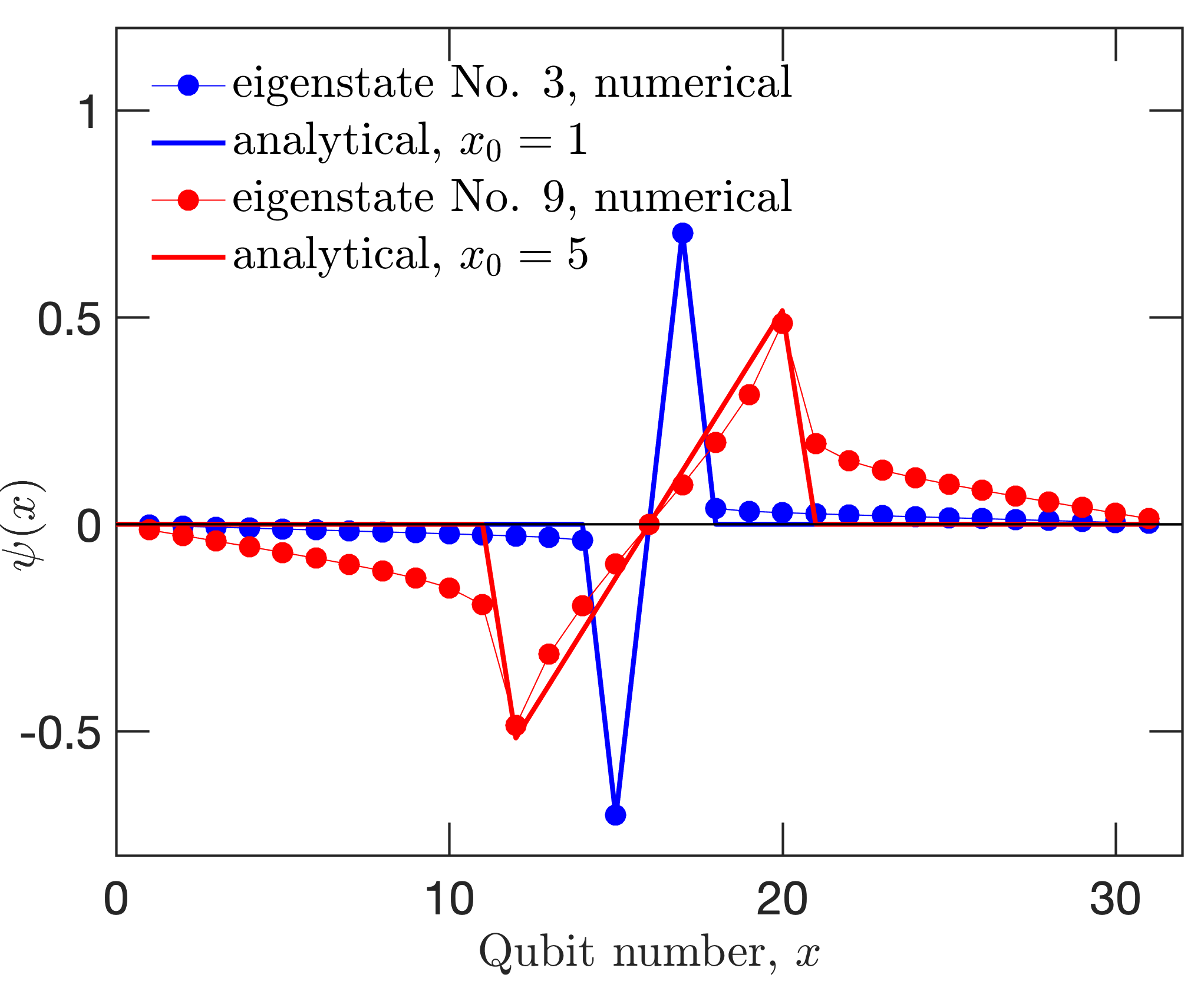}
 \caption{Numerically   calculated solutions of Eq.~\eqref{eq:E1} compared with the analytical answer Eq.~\eqref{eq:anpsi}. Calculation has been performed for the array with $N=31$ qubits and $\kappa=0.1$.}\label{fig:anpsi}
 \end{figure}
In this section we present more detailed analysis of the eigenstates of the  system Eq.~\eqref{eq:E}, \eqref{eq:L},
shown in Fig.~\ref{fig:L} in the main text. For convenience, these equations are repeated below,
\begin{align}
\delta\eps\mathcal E(x,x)&=\mathcal L_{x,x'}\mathcal E(x',x')\:,\label{eq:E1}\\\nonumber
\mathcal L&= 2a\diag[u _0(x)] [\partial_x^{2}-\kappa^2 ]^{-1}\diag[u _0(x)]\partial_{x}^{2}\:.
 \end{align}
 We are interested in the behaviour of solutions in the vicinity of the node of the standing wave, where $u _0(x)=0$. The function 
  $u _0(x)$ can be then expanded up to the linear terms as $u_0(x)= k_{n_0}x$. We have shifted the coordinate origin to the node of $u_0(x)$. We also take into account  that $\kappa \ll 1$.  Equation~\eqref{eq:E1} then transforms to 
  \begin{equation}
  \delta\eps\mathcal E=2\Gamma_0 x\partial_x^{-2}x \partial_x^{2} \mathcal E\:.\label{eq:E2}
  \end{equation}
  Equation \eqref{eq:E2} has mirror symmetry with respect to the point $x=0$. We are interested only in the odd solutions. It can be straightforwardly  shown, that in the continuum approximation Eq.~\eqref{eq:E2} has continuous spectrum  
$\delta\eps=2\Gamma_0 x_0^2$ with the odd eigenstates
  \begin{equation}
  \mathcal E_{\rm odd}(x)=x\theta(x_0^2-x^2)\:.\label{eq:anpsi}
    \end{equation}
    In Fig.~\ref{fig:anpsi} we demonstrate, that  numerical solutions of discrete Eq.~\eqref{eq:E1} are well approximated by the analytical result Eq.~\eqref{eq:anpsi}. The only qualitative difference of the  numerical results from the  analytical answer is the presence of small non-zero background for $|x|>x_0$, that is sensitive to the value of the cutoff parameter $\kappa$. In case of small $x_0$, the solution Eq.~\eqref{eq:anpsi} is strongly localized and looks like a discrete derivative of the discrete $\delta$-function. It can be also checked explicitly that in the continuum regime Eq.~\eqref{eq:E2} has an odd eigenstate $\mathcal E(x)=\delta'(x)$ with $\delta\eps=0$.

\end{document}